\documentstyle[12pt,psfig,paspconf]{article}
\begin{document}

%% Write the title of your paper between the brackets:
{\large\bf{Multiwavelength optical observations of
the chromospherically active binary system MS Ser}}
%{\large\bf{The two active components of MS Ser}}

%% Here comes the author(s) of the paper, please indicate within $^...$
%% the number which corresponds to the institute of each author.
{\it{ J. Sanz-Forcada$^1$, D. Montes$^{1,2}$, M.J.
Fern\'andez-Figueroa$^1$, E. De Castro$^1$, and M.~Cornide$^1$ }}

%% Here write your institute name(s) and address(es),
%% the number in $^..$ indicates your author number, for example:
$^1$ {Departamento de Astrof\'{\i}sica, Facultad de F\'{\i}sicas, Universidad
Complutense de Madrid, E-28040 Madrid, Spain}\\
$^2$ {The Pennsylvania State University,
Department of Astronomy and Astrophysics,
525 Davey Laboratory, University Park, PA 16802, USA}

\vspace*{0.6cm}

To be published  in ASP Conf. Ser., Solar and Stellar Activity:
Similarities and Differences
(meeting dedicated to Brendan Byrne, Armagh 2-4th September 1998)
C.J. Butler and J.G. Doyle, eds

\vspace*{2.5cm}

%---------------------------

\baselineskip=0.6truecm

\large

\hrule
\vspace{0.2cm}
%-----------------------------------------------------3------
\begin{center}
{\huge\bf Abstract}
\end{center}

We present here a continuation of our ongoing project of
multiwavelength optical observations
aimed at studying
the chromosphere of active binary systems
using the information provided for several optical spectroscopic
features that are formed at different heights in the chromosphere
(Montes et al. 1997, 1998; Sanz-Forcada et al. 1998).

In this contribution we focus our study on the preliminar analysis of
the active binary system MS Ser.
We have taken H$\alpha$ and H$\beta$ spectra in 1995
with the Coud\`e Spectrograph at 2.2m telescope in Calar Alto,
and high resolution SOFIN echelle spectra
(covering H$\alpha$,  H$\beta$,
Na~{\sc i} D$_{1}$, D$_{2}$, He~{\sc i} D$_{3}$, Mg~{\sc i} b triplet,
Ca~{\sc ii} H \& K, and Ca~{\sc ii} $\lambda$8662 lines)
in 1998 with the 2.5 m Nordic Optical Telescope (NOT) in La Palma.
A strong emission in the Ca~{\sc ii} H \& K and Ca~{\sc ii} IRT lines,
coming from the primary component (recently classified as K2IV)
is observed.
One of the Ca~{\sc ii} H \& K spectra (at orbital phase near quadrature)
reveals that the secondary (G8V) also exhibit a small emission.
A near complete and variable filling-in of the H$\alpha$ and H$\beta$
is obtained after the application of the spectral subtraction technique.
We detect also some seasonal variations
between these two observing runs and in comparation with
our previous  Ca~{\sc ii} H \& K observations taken in 1993
(Montes et al. 1995).

\vspace{0.4cm}
\hrule

\newpage
\hrule
\vspace{0.2cm}
%---------------------- Introduction --------------------
\begin{center}
{\huge\bf Introduction}
\end{center}
%\large
%\vspace{0.2cm}

Griffin (1978) first observed the binary nature of MS Ser (HD 143313),
calculating the orbital elements for the system through the velocity
curve plotted for the primary star. These orbital elements are presented in
Table 2. Griffin gave its T$_{0}$ in MJD, and this yielded Strassmeier et al.
(1993) to a bad calculation of the T$_{0}$ in HJD. The correct calculus is
the showed in table 2 (phase 0 means the primary star behind).
%Decir algo de que la K2 no se pudo calcular.
Griffin proposed also K2V/K6V as spectral types of the components, based on
photometric arguments for the secondary star.

Bopp et al. (1981) observed a variable filled-in of the H$\alpha$ line, and
calculated a photometric period of 9.60 days, slightly different
from orbital period. Miller \& Osborn (1996) confirm the value of
the photometric period, and Strassmeier et al. (1990) confirm
the active nature
of MS Ser by observing a strong emission in the Ca~{\sc ii} H \& K
composite spectrum.
Dempsey et al. (1993) observed MS Ser within a wide Ca~IRT exploration in
chromospherically active stars. They noted some filled-in in these lines for
MS Ser, but not a reversal emission.
Barrado et al. (1997) measured the Li~{\sc i} 6707.8 equivalent width,
giving 9 m\AA\  as a corrected value for the primary star.

Alekseev (1999)  made a photometric and polarimetric study from MS Ser,
calculating a spot area of 15\% of the total stellar surface, based in part in
the idea that the primary star of MS Ser is a main sequence star. He also
observed some stational variations.
Finally, Osten \& Saar (1998) revised the
stellar parameters for MS Ser, suggesting with their available data
K2IV/G8V as a better classification.

Following the series of papers devoted to the study of RS CVn stars through
the simultaneous analysis of several optical activity indicators (see Montes
et al. 1997, 1998; Sanz-Forcada et al. 1998),
we present spectroscopic observations for MS Ser taken in different epochs
for the H$\alpha$,  H$\beta$, Na~{\sc i} D$_{1}$, D$_{2}$, He~{\sc i} D$_{3}$,
Ca~{\sc ii} H \& K, and Ca~{\sc ii} IRT lines.

We also revise the luminosity class of the
primary star, based on spectroscopic analysis of some metallic lines, like the
Ti~{\sc i} lines, the Hipparcos data, and the Wilson Bappu effect.

\vspace{0.4cm}
\hrule

\newpage
\hrule
\vspace{0.2cm}
%-----------------------------------------
\begin{center}
{\huge\bf Observations}
\end{center}
%\large
%\vspace{0.2cm}

%----------------------------------------------------------

Spectroscopic observations of MS Ser of several optical chromospheric activity 
indicators have been obtained during two observing runs. The first 
was carried out in 1995 June 12th with the 2.2 m 
telescope at the German Spanish Observatory (CAHA) in Calar Alto
(Almer\'{\i}a, Spain), using a Coud\'{e} spectrograph with the
f/3 camera, CCD RCA \#11
covering two ranges: H$\alpha$ (from 6510 to 6638~\AA), and H$\beta$
(from 4807 to 4926~\AA). A resolution of $\Delta\lambda$ 0.26 was achieved
in both cases.

We have also done two observations 
with the 2.56 m Nordic Optical Telescope (NOT) at the Observatorio del Roque 
de Los Muchachos (La Palma, Spain). On 1998 April 5-14, the 
Soviet Finish High Resolution Echelle Spectrograph
(SOFIN) was used with an echelle grating (79 grooves/mm),
camera Astromed-3200, and a 1152x770
pixels EEV P88200 CCD as detector. The wavelength range covers from
3640 to 10085~\AA\ in a total of 40 echelle orders. The
reciprocal dispersion achieved range from 0.07 to 0.18 \AA/pixel.

%echelle observations of 16 chromospheric active binary systems
%have been obtained with the 2.6 m Nordic Optical Telescope (NOT) located
%at the Observatorio del Roque de Los Muchachos (La Palma, Spain).

The spectra have been extracted using the standard reduction procedures
in the IRAF package (bias subtraction, flat-field division, and optimal
extraction of the spectra). The wavelength calibration was obtained by
taking spectra of a Th-Ar lamp. Finally the spectra have been normalized
by a polynomial fit to the observed continuum.

For the spectral subtraction we use $\alpha$ Boo (K1.5III) and HD 45410 
(K0IV) as standard stars, applying 0.85/0.15 as proportional intensities 
%respectively
for the stars.

In Table~1 we give the observing log. For each observation we list 
date, UT, orbital phase ($\varphi$), and the signal to noise ratio (S/N) 
obtained for each spectral region.

Table~2 shows the stellar parameters for the system. We give the correct
calculus for T$_{conj}$ in HJD (phase 0 means the primary star behind). 
Distance is given by the Hipparcos catalogue (ESA, 1997);
B-V, T$_{conj}$ and P$_{orb}$ are given by Griffin (1978); T$_{sp}$, 
Radios and Vsin {\it i} are given by Osten \& Saar (1998), and V-R is given by 
Alekseev (1999).

\vspace{0.4cm}
\hrule

\newpage
%-----------------------------------------

%*****************************************************************
\begin{table*}
\caption[]{Observing log
\label{tab:obslog}}
\begin{flushleft}
%\scriptsize
\small
\begin{tabular}{lccccccc}
\noalign{\smallskip}
\hline
\noalign{\smallskip}
%Cabecera de la tabla
Date &  U.T. & $\varphi$ &
\multicolumn{5}{c}{S/N} \\
 \cline{4-8}
\noalign{\smallskip}
 & & & Ca~{\sc ii} K & H$\beta$ & Na~{\sc i} D & H$\alpha$ &
Ca~{\sc ii} $\lambda$8662 \\
\noalign{\smallskip}
\hline
\noalign{\smallskip}
1995/06/12 & 23:30 & 0.925 & - & - & - & 92 & - \\
1995/06/12 & 23:55 & 0.927 & - & 86 & - & - & - \\
1998/04/08 & 05:05 & 0.206 & 98 & 76 & 98 & 124 & 170 \\
1998/04/11 & 06:15 & 0.545 & 51 & 34 & 35 & 44 & 40 \\
\noalign{\smallskip}
\hline
\end{tabular}
\end{flushleft}
\end{table*}
%*****************************************************************

%*****************************************************************
\begin{table*}
\caption[]{Stellar parameters
\label{tab:par}}
\begin{flushleft}
%\scriptsize
\small
\begin{tabular}{l l c c c c c l l c }
\noalign{\smallskip}
\hline
\noalign{\smallskip}
%             &    &     &    &    &      &    &  & &  \\
{T$_{\rm sp}$} & {SB} & {R} & {d} & {B-V} & {V-R} & {T$_{\rm conj}$}
& {P$_{\rm orb}$} & {P$_{\rm rot}$} & Vsin{\it i}\\
             &    &  (R )  & (pc)  &  &  & (H.J.D.) & (days) &
 (days) & (km s$^{-1}$)  \\
\noalign{\smallskip}
\hline
\noalign{\smallskip}
 K2IV/G8V & 2 & 3.5/1.0  & 88  &  0.94/1.23
& 0.73 & 2442616.142 & 9.01490 & 9.60 & 15/7 \\
%

%..................................................................
\noalign{\smallskip}
\hline
\noalign{\smallskip}
\end{tabular}

\end{flushleft}
\end{table*}
%******************************************************

\vspace{3.2cm}

$\ $

\newpage
\hrule
\vspace{0.2cm}
%-----------------------------------------
\begin{center}
{\Large\bf Luminosity Class}
\end{center}
%\large
%\vspace{0.2cm}

%----------------------------------------------------------

We have revised the luminosity class of the primary star in MS Ser, through the 
study of the behavior of Ti~{\sc i} and some other metallic lines. 
We have compared some of these lines 
(Ti~{\sc i} 8382, Ti~{\sc i} 6625) with those from $\alpha$ Boo 
(K1~III), and HD 45410 (K0~IV). 
These lines are very sensitive to the luminosity 
class for late type stars (see Montes al. 1998), and so we could 
better classify MS Ser primary star as a IV-III, or perhaps a giant. 
We show in Fig~1 the depth of the Ti~{\sc i} 8382 line (multiplet 33).

According to the data provided by the Hipparcos catalogue (ESA, 1997), 
we can calculate, 
through the distance modulus, the absolute magnitude for MS Ser, which is 3.5. 
This yield us to think that the primary star could be a subgiant star.

We have used our Ca~{\sc ii} K spectra to determine
the absolute visual magnitude, M$_{\rm V}$ of the active component
by application of the Wilson-Bappu effect (Wilson \& Bappu 1957).
The mean emission line width measured in our  Ca~{\sc ii} K spectra is
0.56~\AA, which come out to 40.9 km s$^{-1}$ after the quadratic
correction of the instrumental profile.
With the improved Wilson-Bappu relation of Lutz (1970)
we obtained M$_{\rm V}$ = 3.4 and 
with the relation found for chromospherically active binaries
by Montes et al. (1994) we obtained M$_{\rm V}$ = 4.9.
This value is lower than the  M$_{\rm V}$
that corresponds to a K1~V
(M$_{\rm V}$(T$_{\rm sp}$) = 6.6 from
Landolt-B\"{o}rnstein (Schmidt-Kaler 1982)) but higher than for a K1~III
(M$_{\rm V}$(T$_{\rm sp}$) = 0.6).
This result indicates that the primary component of MS Ser may
be of luminosity class IV or higher.

\vspace{0.4cm}
\hrule

\newpage
\hrule
\vspace{0.2cm}
%-----------------------------------------
\begin{center}
{\Large\bf Chromospherical activity indicators}
\end{center}
%\large
%\vspace{0.2cm}

%----------------------------------------------------------

%...ca..............................
{\large\bf\underline{The Ca~{\sc ii} H $\&$ K and H$\epsilon$ lines}}
%.................................

Strong emission in the Ca~{\sc ii} H \& K lines and
the H$\epsilon$ line also in emission arising from the hot component 
is observed in our previous observations of this system in the 
1993 March at orbital phase 0.10 (Montes et al. 1995). 
In the present observations (1998 April) we have deblended the emission 
arising from both components in the spectrum taken at orbital phase near 
quadrature ($\varphi$ = 0.206).
The stronger emission, centered at the absorption line, arise from the hot
component, which is the component with the larger contribution to the
continuum. 
The red-shifted and less intense emission corresponds to the cool
component (see Fig.~2). 
The computed orbital velocity agrees with the observed shift.
In the $\varphi$ 0.545 observation we
can not separate the contribution from both stars.
The H$\epsilon$ line appears also in emission in both spectra.
The emission intensity observed
in our 1993 and 1998 spectra is larger
than the emission intensity observed in the 1988 spectrum presented 
by Strassmeier et al. (1990). 

%sub...cairt......................
{\large\bf\underline{The Ca~{\sc ii} IRT lines}}
%.................................

A clear emission reversal in observed in the core of the 
Ca~{\sc ii} IRT absorption lines $\lambda$8542 and $\lambda$8662.
(see Fig.~3).
After applying the spectral subtraction technique, we can clearly see
a small emission arising from the secondary component, 
in the spectra near quadrature, 
as in the case of the Ca~{\sc ii} H \& K lines.
This emission reversal observed here clearly contrast with the 
only filled-in in these lines reported by Dempsey et al. (1993).

%.................................
{\large\bf\underline{The H$\alpha$, and H$\beta$ lines}}
%.................................

A near complete filling-in H$\alpha$ and H$\beta$ lines are observed in
the 1995 spectra, and a small absorption is present in 1998.
After applying the spectral subtraction a
clear filled-in in the H$\alpha$ and H$\beta$ lines is observed in
the three spectra (see Fig~4 and 5).
The small emission, arising from the secondary component, that is observed in
the Ca~{\sc ii} H \& K and IRT lines is not detected in these two 
Balmer lines.
Bopp et al. (1981) found H$\alpha$ is weakly present in emission 
and variable in intensity. However, we only found a variable filling-in 
in our spectra a two different epochs.

%................................
{\large\bf\underline{He~{\sc i} D$_{3}$ line}}
%................................

We can distinguish this triplet in absorption from the primary star. This is 
typical in cases of activity in evolved stars (Montes et al. 1997). In our 
observations we have measured 35 m\AA\ in the spectrum at $\varphi$ 0.206, 
and 17 m\AA\ at $\varphi$ 0.545
But the line could be slightly blended with other minor 
lines, this could be the responsible of its variation.

\vspace{0.3cm}
\hrule

\newpage
%.................................

\clearpage

%%+++++++++++++++++++
\begin{figure*}
{\psfig{figure=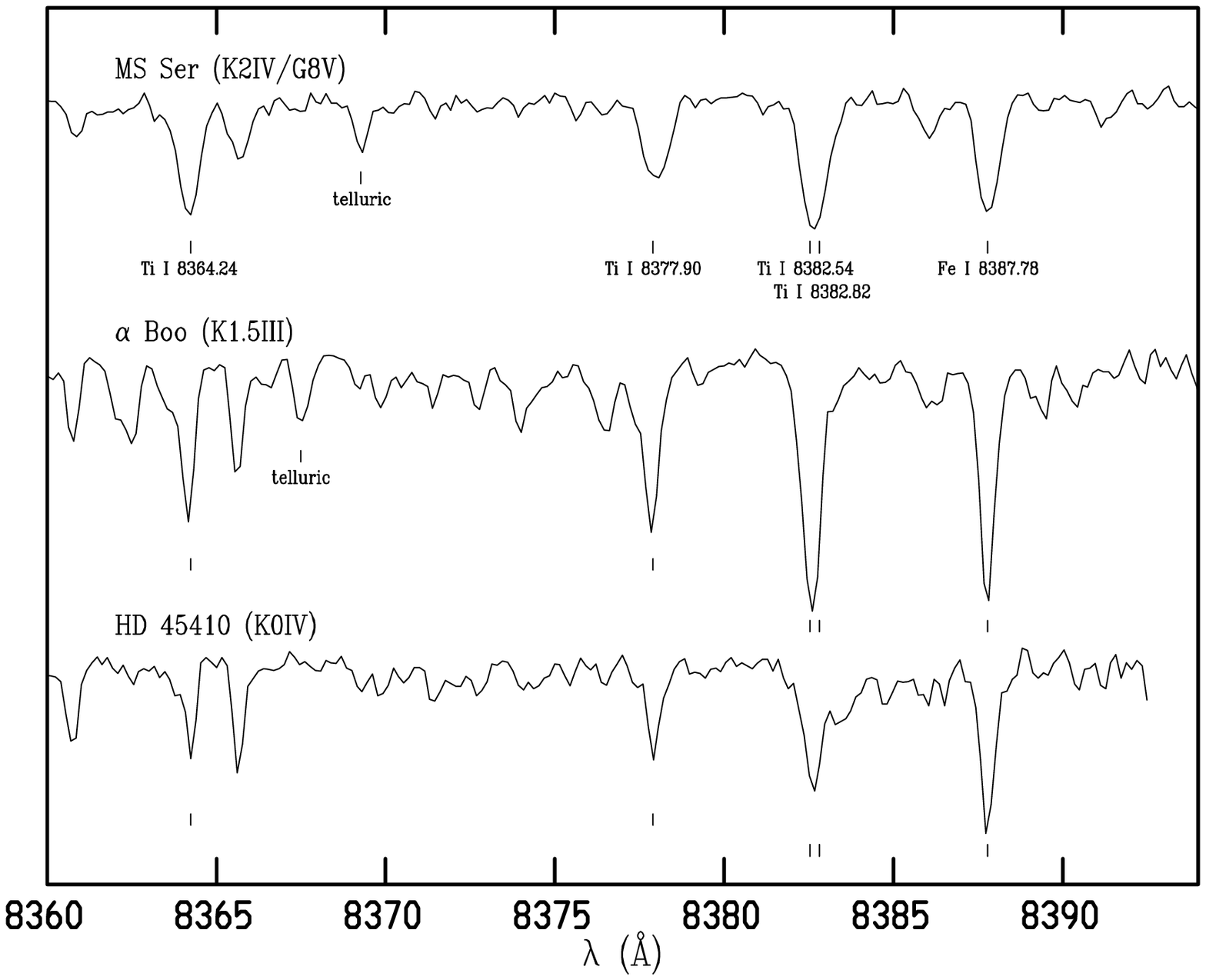,bbllx=94pt,bblly=156pt,bburx=506pt,bbury=653pt,height=19.2cm,width=18.0cm,angle=90}}
\caption[ ]{Spectra of MS Ser and the reference stars 
$\alpha$ Boo (K1~III), and HD 45410 (K0~IV) in the region of several lines of
Ti I (M.33). 
}
\end{figure*}
%%+++++++++++++++++++

%%+++++++++++++++++++
\begin{figure*}
{\psfig{figure=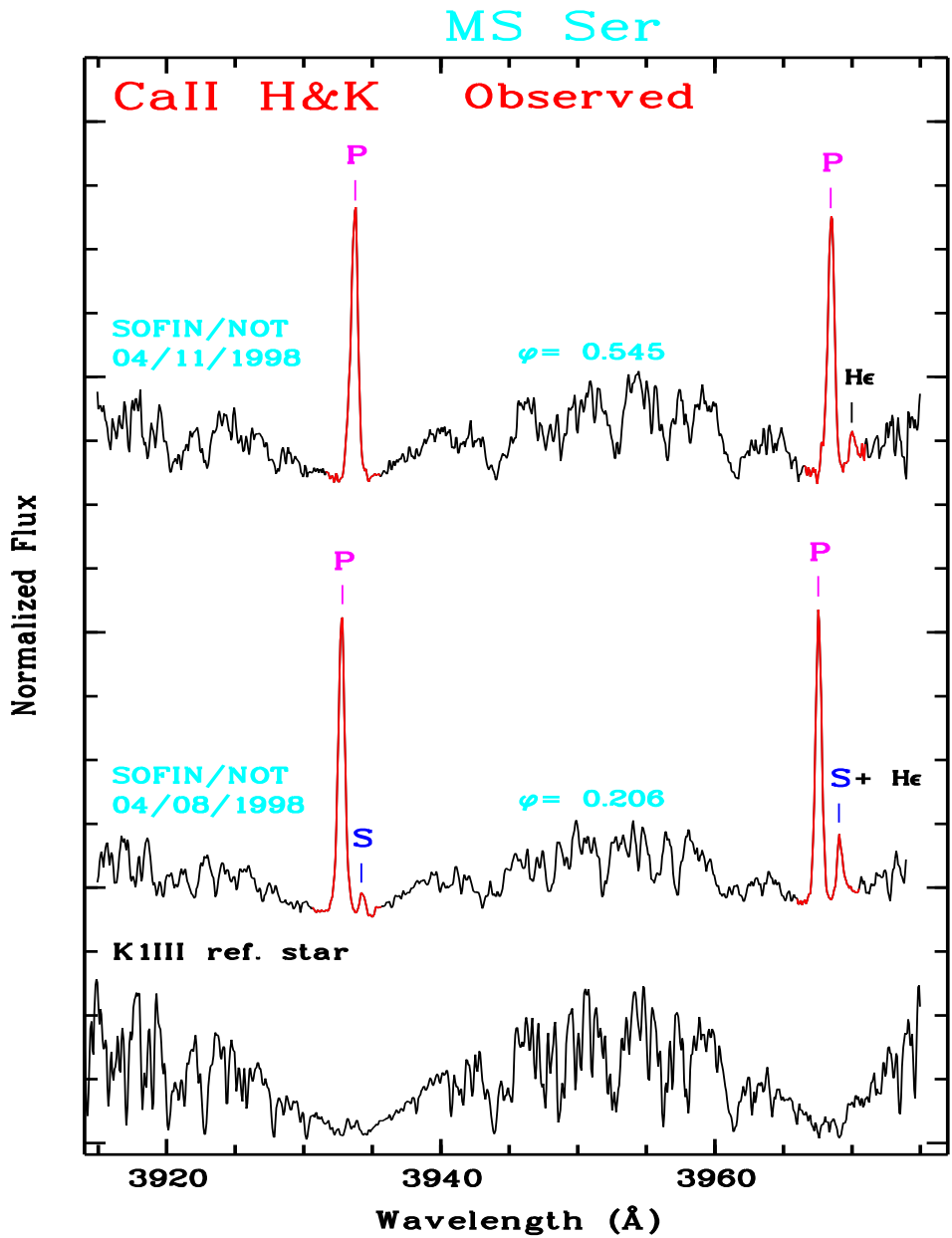,bbllx=45pt,bblly=30pt,bburx=295pt,bbury=462pt,height=16.6cm,width=13.0cm,clip=}}
\caption[ ]{Ca~{\sc ii} H \& K 
observed spectra. The wavelength position of the emission \\
lines of the primary (P) and secondary (S) 
components are marked.
}
\end{figure*}
%%+++++++++++++++++++

%%+++++++++++++++++++
\begin{figure*}
{\psfig{figure=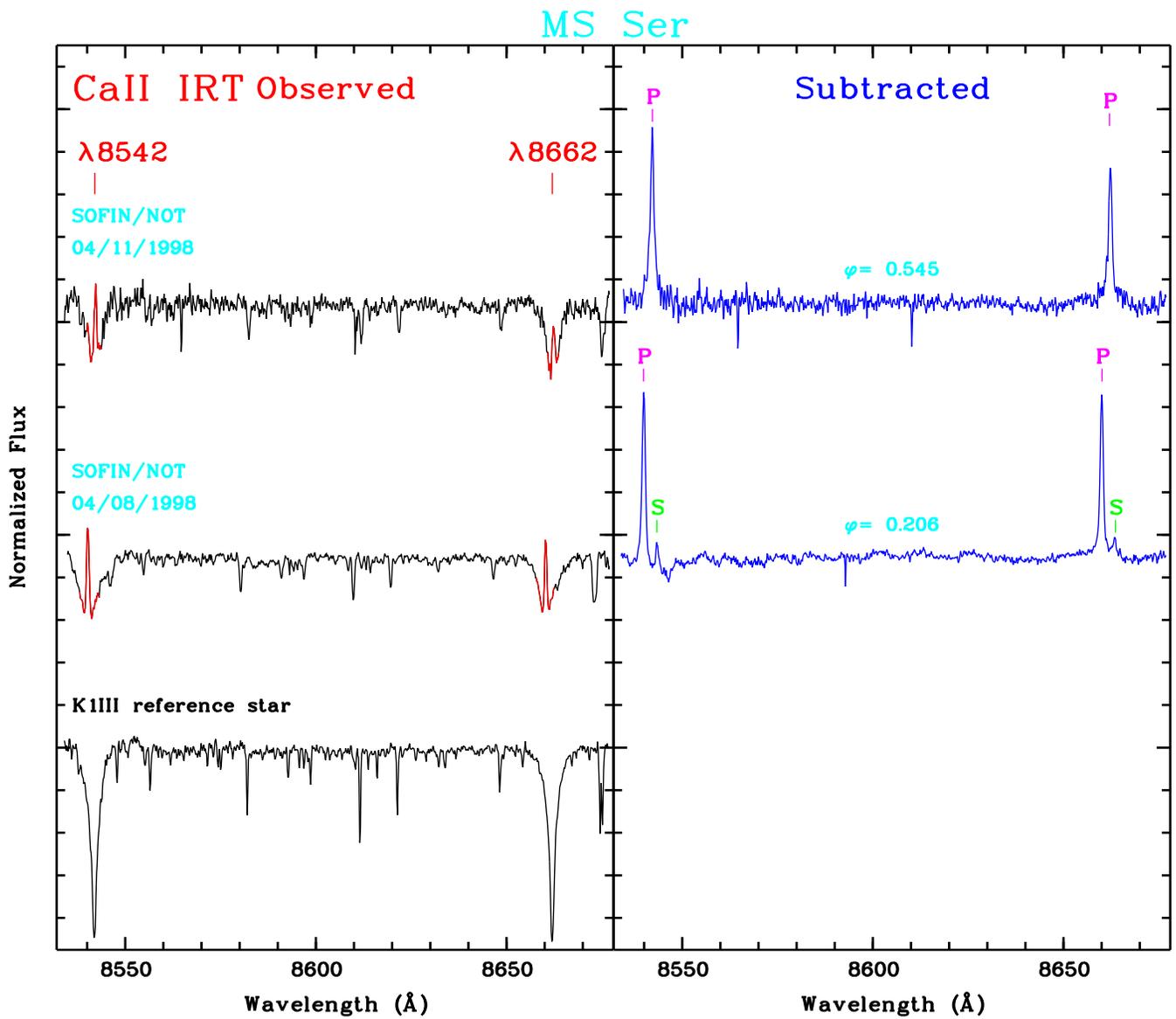,bbllx=45pt,bblly=30pt,bburx=516pt,bbury=462pt
,height=15.6cm,width=18.0cm,clip=}}
\caption[ ]{Ca~{\sc ii} IRT $\lambda$8542 and $\lambda$8662
observed spectra (left panel),
and after the spectral subtraction (right panel).
The wavelength position of 
the emission lines of the primary (P) and secondary (S) components are marked.
}
\end{figure*}
%%+++++++++++++++++++

%%+++++++++++++++++++
\begin{figure*}
\vspace{-0.9cm}
{\psfig{figure=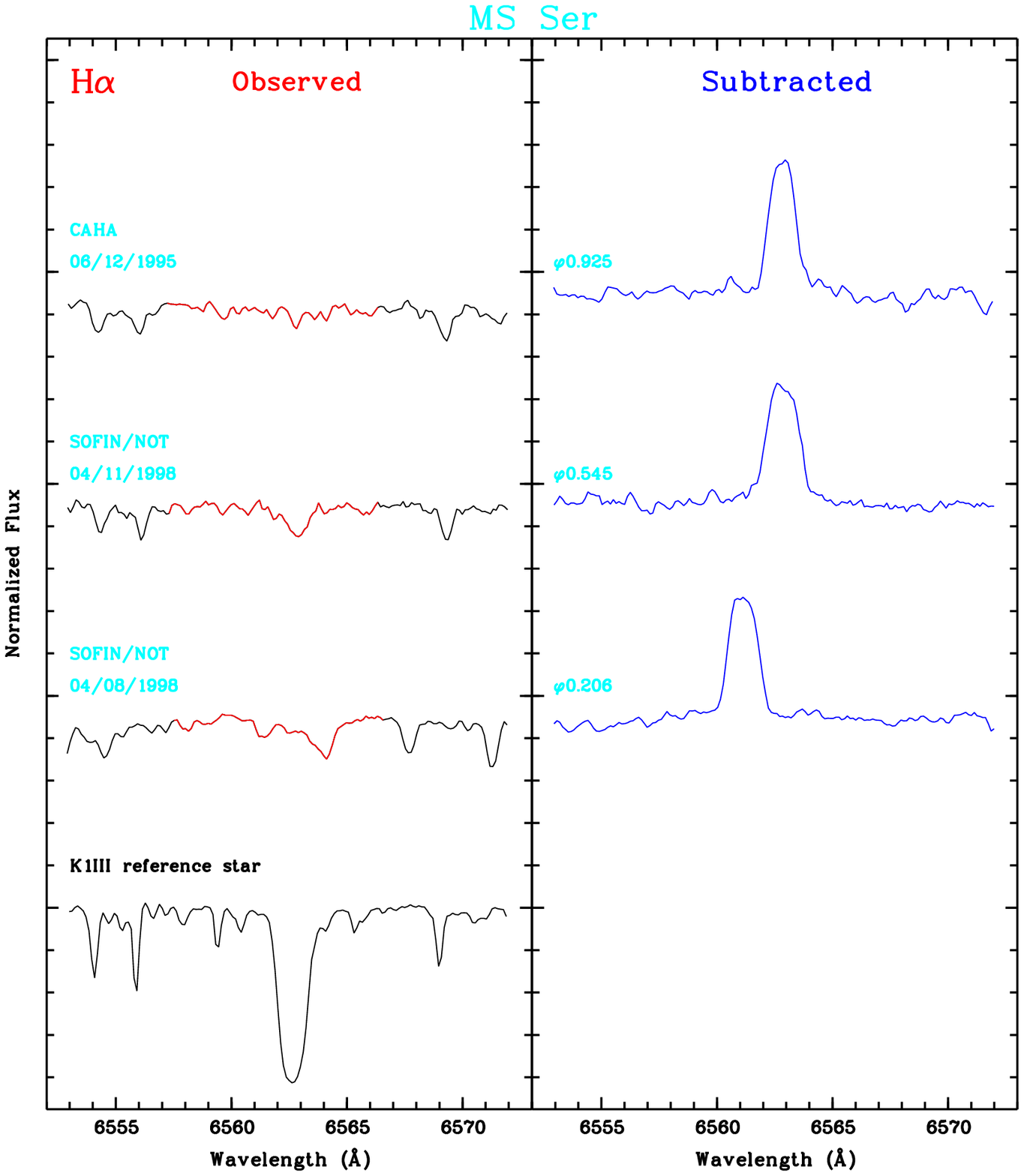,bbllx=45pt,bblly=30pt,bburx=516pt,bbury=569pt,height=20.8cm,width=18.0cm,clip=}}
\caption[ ]{H$\alpha$ observed spectra (left panel),
and after the spectral subtraction (right panel)
}
\end{figure*}
%%+++++++++++++++++++

%%+++++++++++++++++++
\begin{figure*}
\vspace{-0.9cm}
{\psfig{figure=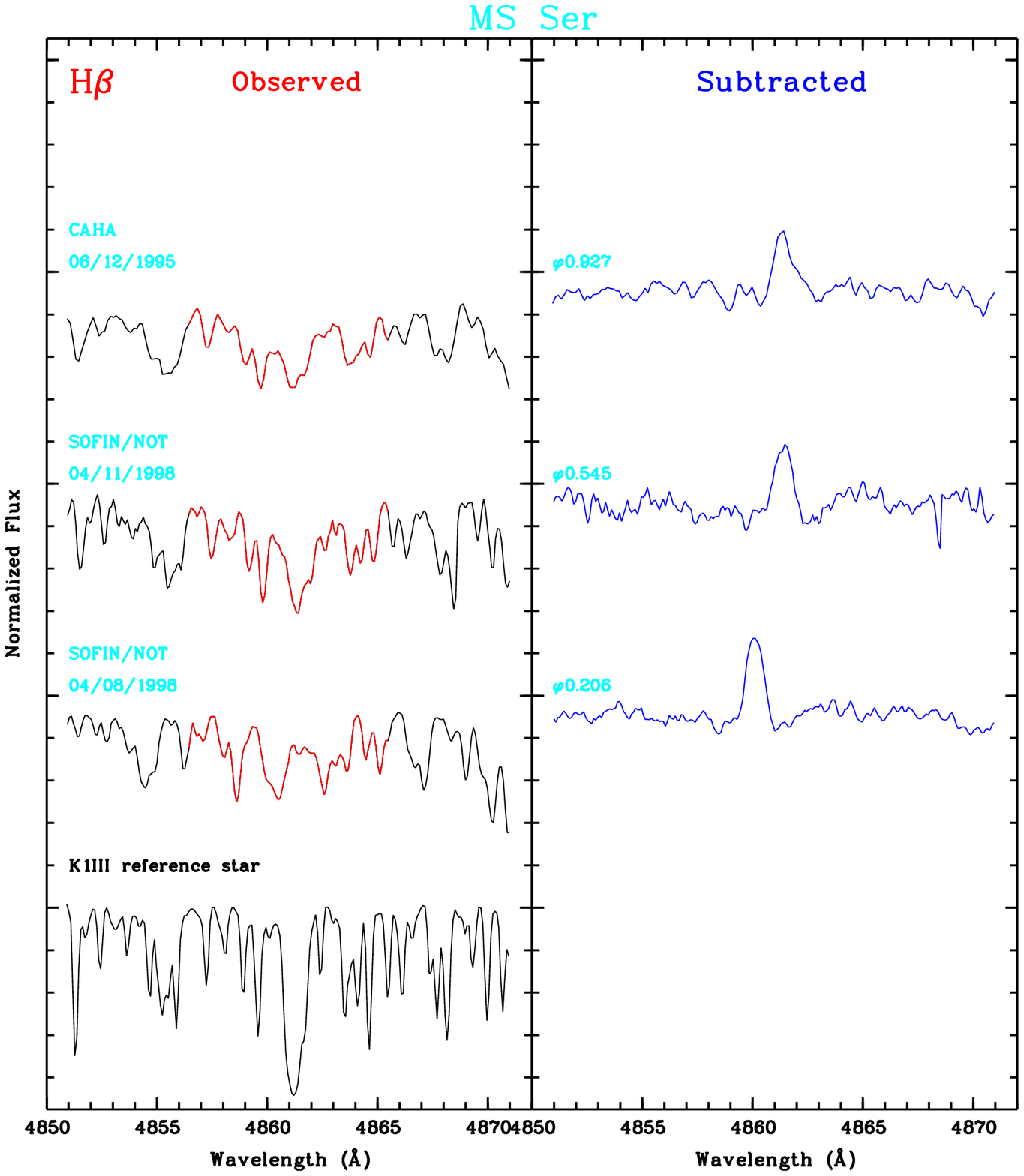,bbllx=45pt,bblly=30pt,bburx=516pt,bbury=569pt,height=20.8cm,width=18.0cm,clip=}}
\caption[ ]{H$\beta$ observed spectra (left panel),
and after the spectral subtraction (right panel)
}
\end{figure*}
%%+++++++++++++++++++

%******************************************************************

\end{document}